\documentclass[letter]{aa}  

\usepackage{graphicx}
\usepackage[varg]{txfonts}
\begin{document}

\authorrunning{Kalari et al.}
   \title{The VLT-FLAMES Tarantula Survey. XV. VFTS\,822: A candidate Herbig B[e] star at low metallicity \thanks{Based on the observations at the European Southern Observatory Very Large Telescope in programme 182.D-0222.}, \thanks{Table 1 and Figure 4 in on\,line version only.}}
\titlerunning{VFTS\,822 a Herbig B[e] star in 30 Doradus?}
   \subtitle{}
   \author{V. M. Kalari\inst{1, 2}, 
J. S. Vink\inst{1}, 
P. L. Dufton \inst{2},
C. J. Evans\inst{3}, 
P. R. Dunstall\inst{2},
H. Sana\inst{4},
J. S. Clark\inst{5},
L. Ellerbroek \inst{6},
A. de Koter \inst{6, 7},
D. J. Lennon \inst{8}, 
W. D. Taylor\inst{3}
          }

   \institute{Armagh Observatory, College Hill,
              Armagh, BT61\,9DG, UK, \email{vek@arm.ac.uk}
   \and
       Department of Physcis \& Astronomy, Queen's University Belfast, BT7\,1NN, UK \
   \and
       UK Astronomy Technology Centre, Royal Observatory, Edinburgh, Blackford Hill, Edinburgh, EH9\,3HJ, UK \
    \and
     ESA/STScI, 3700 San Martin Drive, Baltimore, MD\,21218, USA \ 
    \and
     Department of Physics and Astronomy, The Open University, Walton Hall, Milton Keynes, MK7\,6AA, UK \
    \and
     Astronomical Institute Anton Pannekoek, Amsterdam University, Science Park 904, 1098\,XH, Amsterdam, The Netherlands \
    \and 
     Institue voor Sterrenkunde, KU Leuven, Celestijnenlaan 200D, 3001 Leuven, Belgium \ 
    \and
     European Space Astronomy Centre, Camino bajo del Castillo, Villanueva de la Ca\~{n}ada,   E-28692 Madrid, Spain  
             }

   \date{Received 19 December 2013 / Accepted 13 January 2014}

\abstract {We report the discovery of the B[e] star VFTS 822 in the 30 Doradus star-forming region of the Large Magellanic Cloud, classified by optical spectroscopy from the VLT-FLAMES Tarantula Survey and complementary infrared photometry. VFTS 822 is a relatively low-luminosity (log\,$L$~=~4.04\,$\pm$\,0.25\,$L_{\odot}$) B8[e] star. In this Letter, we evaluate the evolutionary status of VFTS 822 and discuss its candidacy as a Herbig B[e] star. If the object is indeed in the pre-main sequence phase, it would present an exciting opportunity to spectroscopically measure mass accretion rates at low metallicity, to probe the effect of metallicity on accretion rates.}


   \maketitle

%

\setlength{\parskip}{0.1 cm plus0mm minus1.5mm}
\setlength{\textfloatsep}{0.1 cm} 
\section{Introduction}

Present understanding suggests that most stars form by disc accretion during pre-main sequence (PMS) evolution. In the favoured accretion model, the disc forms during protostellar collapse and is disrupted by a stellar magnetosphere, which channels gas from the disc onto the stellar surface (see Hartmann 2008). Practical limitations have restricted observations of PMS stars to nearby solar-metallicity star-forming regions. Consequently, the properties of low-metallicity PMS stars, located in extra-galactic star-forming regions remain largely unknown.

The formation of stars may be significantly influenced by the metallicity ({\it Z}) of the local medium. The extra cooling provided by increasing {\it Z} is thought to decrease stellar mass. This is perhaps exemplified in the contrast between the present-day low-mass stellar initial mass function (IMF) and the anticipated high-mass population III IMF (Larson 1998). This may imply higher mass accretion rates at lower {\it Z} (Omukai \& Palla 2001).

The only current possibility of observing low {\it Z} PMS stars spectroscopically is in the Magellanic clouds. Medium-resolution spectra, rather than photometric observations, are necessary as high-mass PMS systems at large distances become increasingly confused. They cannot be separated from more evolved systems occupying similar regions of the Hertzsprung-Russell (H-R)  diagram using photometry alone. Another driver for observing PMS stars spectroscopically is the possibility to measure mass accretion rates using different indicators. 

Medium-resolution stellar spectroscopy with eight-to ten-metre class telescopes leads to a limiting magnitude of {\it V}~$\sim$~18.
In the nearest low-{\it Z} galaxy, the Large Magellanic Cloud (LMC) located at a distance of 50\,kpc (Schaefer 2008), this translates to a
mid-late B spectral type. These stars are more massive than 4\,$M_{\odot}$ and reside near the main sequence with a very
short PMS phase. Therefore, detection of even a single PMS star 
is contentious, although several candidates have been reported (de Wit et al. 2005; Clayton et al. 2010; De Marchi et al. 2010).

One approach to identifying high-mass PMS stars is to discover them serendipitously from large spectroscopic surveys. The VLT-FLAMES Tarantula survey (VFTS; Evans et al. 2011) covering over 800 massive stars in the 30 Doradus starburst region of the LMC ($Z$~$\sim$~$0.5\,Z_{\odot}$) provides an ideal dataset for this task. From the VFTS sample, we identify VFTS\,822 (RA(J2000)\,=\,05$^{\text{h}}$39$^{\text{m}}$38.49$^{\text{s}}$, Dec(J2000)\,=\, $-$69$^{\circ}$09\arcmin00.5\arcsec) as an accreting PMS candidate. We describe the characteristics of VFTS\,822 that indicate it is in an accreting phase. However, since VFTS\,822 exhibits the B[e] phenomenon (Lamers et al. 1998), its evolutionary status is unclear. Therefore, we explore the possibility that it could also be a B[e] supergiant.

\section{Results}
  \subsection{VFTS spectroscopy}

\begin{figure*}[t]
\centering   
\label{Fig. 1}
 \includegraphics[width=150mm, height=105 mm]{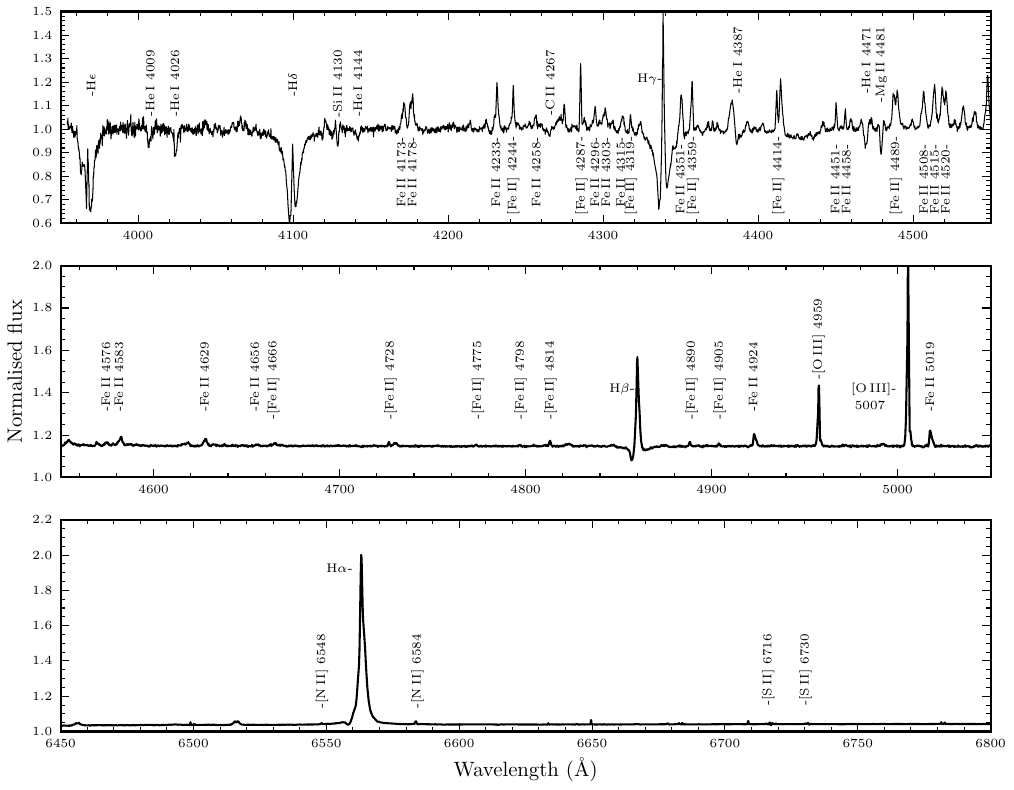}
 \caption{Median FLAMES-Giraffe spectra of VFTS\,822.}
 \label{}
\end{figure*} 

Multi-epoch observations covering 3950\,-\,5050\,\AA~($R$\,$\approx$\,8\,000) and 6450\,-\,6800\,\AA~($R$\,$=$\,16\,000) were taken using the FLAMES multi-fibre spectrograph (Pasquini et al. 2002). Further details on the observations and data reduction techniques can be found in Evans et al. (2011). Spectra of VFTS\,822 are plotted in Fig.~1. Strong H$\alpha$, H$\beta$, and H$\gamma$ emission are seen, together with emission lines from both permitted and forbidden metal transitions. The intensity of the Si\,{\scriptsize II} $\lambda$4128-30 and Mg\,{\scriptsize II} $\lambda$4481 absorption, combined with the relative weakness of the He\,{\scriptsize I} lines at, e.g., $\lambda\lambda$4009, 4026, argue for an approximate classification of B8 (particularly since we expect that the He\,{\scriptsize I} $\lambda$4471 line is partly filled in by nebular contamination). Radial velocity ($v_{\text {rad}}$) estimates were obtained by profile fitting the He\,{\scriptsize I} lines at $\lambda\lambda$4009, 4026 and 4143 and the Mg\,{\scriptsize II} $\lambda$4481 doublet. A mean value of 262\,$\pm$\,7\,km\,s$^{-1}$ was obtained. No evidence for binarity was found from peak-to-peak $v_{\text {rad}}$ variations. 

The estimation of atmospheric parameters is difficult because of the emission line spectrum that implies that a non-LTE photospheric model may not be appropriate. However, we have tried to provide constraints using the {\sc tlusty} model atmospheres grid (Lanz \& Hubeny 2007). The H$\delta$ and H$\gamma$ profiles provide loci of estimates in the effective temperature and logarithmic gravity plane ($T_{\text {eff}}$, log\,{\it g}). Possible solutions include (12000\,K, 2.6) and (15000\,K, 2.9). Assuming a normal helium abundance, the profile of the He\,{\scriptsize I} $\lambda$4026 line is consistent with the former. The only well-observed metal line is the Mg\,{\scriptsize II} doublet, which leads to an upper magnesium abundance estimate (assuming zero microturbulence) of 6.87 dex for the lower $T_{\text {eff}}$. This is slightly smaller than found by Hunter et al. (2007) for the LMC using the same model atmosphere grid. As a result, $T_{\text {eff}}$ is unlikely to be significantly lower than 12000\,K. We adopt $T_{\text {eff}}$~=~12000\,$\pm$\,3000\,K and log\,{\it g}~=~2.6\,$\pm$\,0.4) as our best estimates. For spectra of this quality, the errors would normally be of the order of ($\pm$\,1000\,K, $\pm$\,0.1), but the uncertainties in the appropriateness of the modelling implies that there might be additional uncertainties. 

The H$\alpha$ equivalent width (EW$_{\text{H}\alpha}$)~=~$-$67.5\,$\pm$\,4\,$\AA$. The error reflects the difference obtained by fitting either the wings or the core. The EW$_{\text{H}\alpha}$ is greater than maximum measured values for classical Be stars of $\gtrsim$ ~$-$50$\AA$ (Jones et al. 2011). The H$\beta$ equivalent width (EW$_{\text{H}\beta}$)~=~$-$2.72\,$\pm$\,0.5\,\AA.

\subsection {Photometry} 

\onltab{
\begin{table}
\caption{Adopted photometry for VFTS\,822}
\label{table. 1}
\centering
\resizebox{\columnwidth}{!}{%
\begin{tabular}{|c|c|c|c|c|c|}
\hline
  \multicolumn{1}{|c|}{Band} & 
  \multicolumn{1}{c|}{Magnitude} &
  \multicolumn{1}{c|}{Ref.} & 
  \multicolumn{1}{c|}{Band} &
  \multicolumn{1}{c|}{Magnitude} &  
  \multicolumn{1}{c|}{Ref.}\\
\hline
  {\it U} & 15.395\,$\pm$\,0.03 & 1 & 3.3\,$\mu$m & 10.648\,$\pm$\,0.02 & 3\\
  {\it B} & 16.028\,$\pm$\,0.31 & 1 & 3.6\,$\mu$m & 10.402\,$\pm$\,0.02 & 4\\
  {\it V} & 15.433\,$\pm$\,0.02 & 1 & 4.5\,$\mu$m & 9.775\,$\pm$\,0.02 & 4\\
  {\it I} & 14.926\,$\pm$\,0.02 & 1 & 4.6\,$\mu$m & 9.696\,$\pm$\,0.02 & 3\\
  {\it J} & 14.246\,$\pm$\,0.03 & 2 & 5.8\,$\mu$m & 8.206\,$\pm$\,0.02 & 4\\
  {\it H} & 13.557\,$\pm$\,0.02 & 2 & 8.0\,$\mu$m & 8.662\,$\pm$\,0.02 & 4\\
  {\it K}s & 12.296\,$\pm$\,0.02 & 2 & 11.6\,$\mu$m & 7.383\,$\pm$\,0.02 & 3\\
          &                &                & 22.1\,$\mu$m & 3.606\,$\pm$\,0.04 & 3\\
\hline
\end{tabular}}
\tablebib{(1) Zaritsky et al. (2004); (2) Cutri et al. (2003); (3) Cutri et al. (2012); (4) Meixner et al. (2006) }
\end{table}}

Broad-band photometry for VFTS\,822 was adopted from the published catalogues summarised in Table~1. The broadband photometry was not used to estimate extinction, because the emission line contribution to the continuum, although small, cannot be well constrained. We therefore employed equivalent width measurements of the $\lambda$4428 and $\lambda$6614  diffuse interstellar band features. From van Loon et al. (2013), $A_{\text V}$~=~EW$_{4428}$/\,0.7 and $A_{\text V}$~=~EW$_{6614}$/\,0.03. This leads to an estimated extinction of $A_{\text V}$~=~1.75\,$\pm$\,0.05 (where the observed scatter is not accounted for). 

Using an ATLAS9 model (Castelli \& Kurucz 2004) having $T_{\text {eff}}$~=~12000\,K, log\,{\it g}~=~2.6, [M/H]=-0.5, we calculated a bolometric correction, BC~=~$-$0.673. We estimate the logarithm of the luminosity, log\,$L$~=~4.04\,$\pm$\,0.25\,$L_{\odot}$. The error accounts for the variation in BC due to the uncertainty in $T_{\text {eff}}$. This is most likely an upper limit, since the disc contribution is considered negligible. The luminosity is greater than expected for a mid-late B-type star, but not unprecedented for Herbig stars. For example, HD\,85567, V594~Cas, and V921~Sco each have log\,$L$~$\gtrsim$~4\,$L_{\odot}$ and are all classified as Herbig B[e] stars (Lamers et al. 1998; Kraus et al 2012).

\subsection {Infrared properties}

\begin{figure}[t]
\label {fig. 2}
\resizebox{\hsize}{!}{\includegraphics[width=15 mm, height=10 mm]{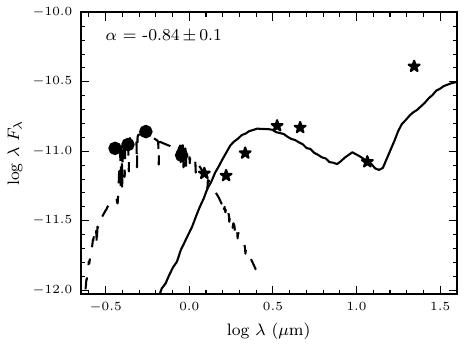}}
\caption{Observed fluxes in optical (circles) and IR (asterisks). The dashed line is the model SED and the solid line the best-fit YSO model.}
\label{}
\end{figure} 

The SED slope ($\alpha$) in the infrared is an essential tool for diagnosing the nature of any infrared excess. We follow Lada et al. (1987), where $\alpha$~=~{\it d}\,log($\lambda$$F_{\lambda}$)/\,{\it d}\,log$\lambda$ and $\lambda$~>~3\,$\mu$m. WISE photometry was not used to calculate $\alpha$ because it is possibly affected by nebulosity at $\lambda$~>~10\,$\mu$m. We therefore calculated $\alpha$ based on {\it Spitzer} photometry alone. For VFTS\,822, $\alpha$~=~$-$0.84\,$\pm$\,0.1, which is greater than $\alpha$~$\sim$~$-$3, as expected for normal OB stars (Sung et al. 2009). It is a Class II PMS object based on the scheme devised by Lada et al. (1987). The SED slope closely resembles the one for objects with a dusty disc, $\lambda$$F_{\lambda}$~$\propto$~$\lambda^{-2/3}$, i.e., a much slower decrease with wavelength than for most evolved stars. 

A diagnostic feature of dusty discs is ({\it K}\,$-$\,24\,${\mu}$m)~$\gtrsim$~5 (e.g. Hernandez et al. 2006). For VFTS\,822, {\it K}s\,$-$\,22\,${\mu}$m~$\sim$~9\,mag, indicating the likely presence of a dusty disc. In Fig. 2, we plot the optical-infrared SED for VFTS\,822. Overplotted is the extinction corrected adopted ATLAS9 model spectrum. The solid line is the best-fit young stellar object (YSO) model ($\chi^{2}$~=~85) from Robitalle et al. (2007). The SED slope and infrared colours indicate that VFTS\,822 has a dusty circumstellar disc characterised by a disc temperature of $\sim$950\,K. 

\section{Evolutionary status}
In low-mass stars, emission line spectra and dust discs are unique to the PMS phase. However, at Herbig Be masses, B[e]-type stars also display similar characteristics. B[e]-type stars are rare and heterogeneous in terms of their evolutionary status. They include both PMS objects and evolved supergiants (sg). Based on the spectral type, emission line features, and infrared properties, we consider VFTS\,822 to be a B[e]-type star (see Lamers et al. 1998). Identifying the evolutionary status of VFTS\,822 is challenging because the observed properties of sgB[e]-type stars and PMS Herbig-B[e] stars that have {\it L}~$\sim$~4-4.5\,$L_{\odot}$ are similar. Therefore, in the following sections we discuss the possibility that VFTS\,822 might be either a PMS B[e] star or an sgB[e] star. 

\subsection{B[e] supergiant?} 

The best fit log\,{\it g} (2.6\,$\pm$\,0.4) is larger than the values measured for many late B-type sg (cf. Firnstein \& Przybilla 2012, where log\,{\it g}~$\sim$~2 is typical). Owing to the emission lines spectra, abundances are challenging to determine, but we do not see the compelling evidence of chemical evolution that we might expect for an sgB[e] star (e.g. no N\,{\scriptsize II} $\lambda\lambda$3995, 4630 features are seen). However, the lack of evidence for He enrichment suggests that nitrogen enrichment need not necessarily be present. Therefore, one cannot exclude the possibility that VFTS\,822 is a supergiant. One promising avenue forward would be to compare the enrichment of CO$^{13}$ to CO$^{12}$ infrared bands (Kraus et al. 2009).

If VFTS\,822 is an sgB[e] star, it has an age of 36\,Myr, with a lower error bar of 4\,Myr based on the single star isochrones of Marigo et al. (2008). This is significantly older than the estimated age of R136, the central cluster of 30~Dor (1-2\,Myr), and the high-mass population in the surrounding nebula (e.g. Walborn \& Blades, 1997). The $v_{\text {rad}}$ of VFTS\,822 is similar to other stars in the region (Sana et al. 2013), making it unlikely to be an interloper. Nonetheless, the possibility of a genuine 36\,Myr old sgB[e] cannot be ruled out because of the complex star-formation scenario and age spread present within 30~Dor. However, the inherent rarity of sgB[e] stars means that a large initial population is expected to yield them (Clark et al. 2013). Such a population, which should be readily identifiable via the resulting IR bright red supergiant/super-AGB cohort, is not known to be present within the immediate vicinity (Bonanos et al. 2009).

\subsection{Or Herbig B[e] star?}

The position of VFTS\,822 in the H-R diagram is compared to known LMC sgB[e] (Zickgraf 2006) and Galactic Herbig B[e] stars (Lamers et al. 1998) in Fig~3. No Herbig B[e] stars have been discovered yet in the LMC. Birthlines with mass accretion rates ($\dot{M}_{\text{acc}}$) of 10$^{-4}$ and 10$^{-3}$\,$M_{\odot}$$yr^{-1}$ from Hosokawa \& Omukai (2009) are plotted, along with the zero-age main sequence from Schaerer et al. (1992) and 10$^{4.5}$\,yr PMS isochrone from Bressan et al. (2012). The position of VFTS\,822 on the H-R diagram is coincident with a low-luminosity LMC sgB[e], but it 
sits in an H-R diagram position that is also covered by known Galactic Herbig B[e] stars. The comparable positions of sgB[e] and Herbig stars at log\,$L$~<~4.5\,$L_{\odot}$ make establishing the evolutionary status of VFTS\,822 non-trivial.

If VFTS\,822 is a PMS, it has a $\dot{M}_{\text{acc}}$ of 3\,$\pm$\,1\,$\times$\,10$^{-4}$\,$M_{\odot}$$yr^{-1}$, a mass ({\it M}$_{*}$) of 10.7\,$\pm$\,1\,$M_{\odot}$, and age $\simeq$ 30000\,yrs according to the PMS models. PMS stars at such masses have extremely short PMS lifetimes, but a few have been detected (Drew et al. 1997). VFTS\,822 is expected to be past its protostellar phase and to be accreting mass. At this age and mass, the $\dot{M}_{\text{acc}}$ of the object is expected to be relatively high, as is necessary to overcome the radiative pressure. Hosokawa \& Omukai (2009) predict that at high $\dot{M}_{\text{acc}}$ (>~10$^{-4}$\,$M_{\odot}$$yr^{-1}$), massive stars (>~10\,$M_{\odot}$) are seen in the PMS phase. We find that the theoretically predicted $\dot{M}_{\text{acc}}$ are similar to the observed accretion rates calculated in Sec. 3.2.1.

The estimated log\,{\it g} of VFTS 822 (2.6\,$\pm$\,0.4) is somewhat lower than expected for PMS stars. However profile fitting of Herbig stars for log\,{\it g} determination is not common practice, with a recent result showing a Galactic Herbig star to have a log\,{\it g} value of 3.5, similar to giant stars (Ochsendorf et al. 2011). Given the modest extinction toward VFTS 822, this implies a relatively large stellar radius of over 10\,$R_{\odot}$. This may, however, be symptomatic of a bloated star, where very high $\dot{M}_{\text{acc}}$ causes stars to bloat

VFTS\,822 is located near the edge of the 30\,Dor nebula at 5.8\arcmin~from R136 (approx. 84.4\,pc at the distance to the LMC) as shown in Fig.~4. This region contains previously identified loosely grouped Class II PMS objects, which includes the Spitzer counterpart of VFTS\,822 (Kim et al. 2007), whereas YSOs or Class I objects are situated nearer to R136. It is located on the edge of filamentary structure, where one might expect to find Class II PMS objects (Hartmann 2008).

\subsubsection{$\dot{M}_{\text{acc}}$}
Balmer line emission, especially H$\alpha$, is a commonly used accretion tracer. It is thought to originate in the magnetospheric accretion columns (Calvet \& Hartmann 1992). The line luminosity ({\it L}$_{\text {line}}$) can be used to measure the $\dot{M}_{\text{acc}}$. To calculate {\it L}$_{\text {line}}$ from H$\alpha$ and H$\beta$, we assume that the excess flux in the continuum is negligible. This assumption is correct if there is no significant veiling. Since photospheric absorption lines are visible, we assume this is justified. The line flux ($F_{\text {line}}$) is given by 
\begin{equation}
    \ F_{\text {line}}= F_{\text {cont}} \times \text{EW}_{\text {line}}.\
\end{equation} 
Here $F_{\text {cont}}$ is the continuum flux. The continuum flux was measured using the adopted ATLAS9 model spectra. The average flux of the model over the 6556\,-\,6570\,\AA and 4856\,-\,4864\,$\AA$ ranges was taken as the H$\alpha$ and H$\beta$ continuum flux, respectively. Multiplying by 4$\pi$$d^{2}$, we obtain the respective {\it L}$_{\text {line}}$. Based on the accretion luminosity ({\it L}$_{\text {acc}}$)-{\it L}$_{\text {line}}$ relation of Herczeg \& Hillenbrand (2008), we determine log\,{\it L}$_{\text {acc}}$~=~3.89\,$\pm$\,0.6\,$L_{\odot}$ from H$\alpha$ and log\,{\it L}$_{\text {acc}}$~=~3.58\,$\pm$\,0.6\,$L_{\odot}$ from H$\beta$. The errors include the RMS scatter in the {\it L}$_{\text {acc}}$-{\it L}$_{\text {line}}$ relations.  

Another commonly used accretion indicator is the Balmer jump. The excess luminosity is thought to be caused by an accretion shock on the stellar surface. The UV excess of VFTS\,822, ({\it U$-$I})\,-\,({\it U$-$I})$_{0}$~=~$-$0.742, can be used to measure its accretion rate. The observed ({\it U$-$I}) colour was corrected for extinction, and the ({\it U-I})$_{0}$ model colour was determined using the model spectrum and appropriate filter responses. The excess U-band luminosity, log\,$L_{(\text{U, excess})}$~=~2.96\,$\pm$\,0.3\,$L_{\odot}$, and the accretion luminosity {\it L}$_{\text {acc}}$\,=\,3.96\,$\pm$\,0.5\,$L_{\odot}$. The $\dot{M}_{\text{acc}}$ is given by the free-fall equation. Following the assumptions of Herczeg \& Hillenbrand (2008), 
\begin{equation}
    \ \text{log}\,\dot{M}_{\text {acc}}= -7.39+ \text{log}\,L_{\text {acc}}/L_{\odot} + \text{log}\,R_{*}/R_{\odot}-\text{log}\,M_{*}/M_{\odot}.
\end{equation}
Here $\dot{M}_{\text{acc}}$ is in units of $M_{\odot}$$yr^{-1}$. We calculate a median $\dot{M}_{\text{acc}}$ of 3.16\,$\pm$\,2\,$\times$\,10$^{-4}$\,$M_{\odot}$$yr^{-1}$ from the three indicators. The {\it L}$_{\text {acc}}$-{\it L}$_{\text {line}}$, {\it L}$_{\text {acc}}$-{\it L}$_{\text {U,excess}}$ relations adopted are tested well on {\it Z}$_{\odot}$ PMS stars, up to Herbig Ae masses ($\lesssim$~4\,$M_{\odot}$). We therefore acknowledge that they may not be numerically applicable to VFTS\,822, but only provide an order-of-magnitude estimate. The applied magnetospheric accretion model has been considered valid for spectral types up to late B, but not for the majority of Herbig Be stars (Mottram et al. 2007; Monnier et al. 2005).

The agreement between the $\dot{M}_{\text{acc}}$ calculated from the H$\alpha$, H$\beta$, {\it U}-band luminosities suggests that the Balmer lines and jump are produced in the same mass accretion phenomenon, favouring a PMS evolutionary phase. 

\begin{figure}[t]  
\label {fig. 3}
\resizebox{\hsize}{!}{\includegraphics[width=50 mm, height=37.5 mm]{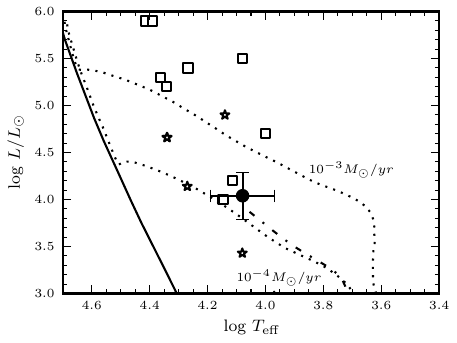}}
\caption{H-R diagram showing the position of VFTS\,822 (circle), LMC sgB[e] (squares), Galactic PMS B[e] (asterisks), model birthlines (dotted lines), MS (solid line), and PMS (dashed-dotted line) isochrones.} 
\label{}
\end{figure} 

\onlfig{
\begin{figure}[t]
\label {fig. 4}
\resizebox{\hsize}{!}{\includegraphics[width=5 mm, height=4.1 mm]{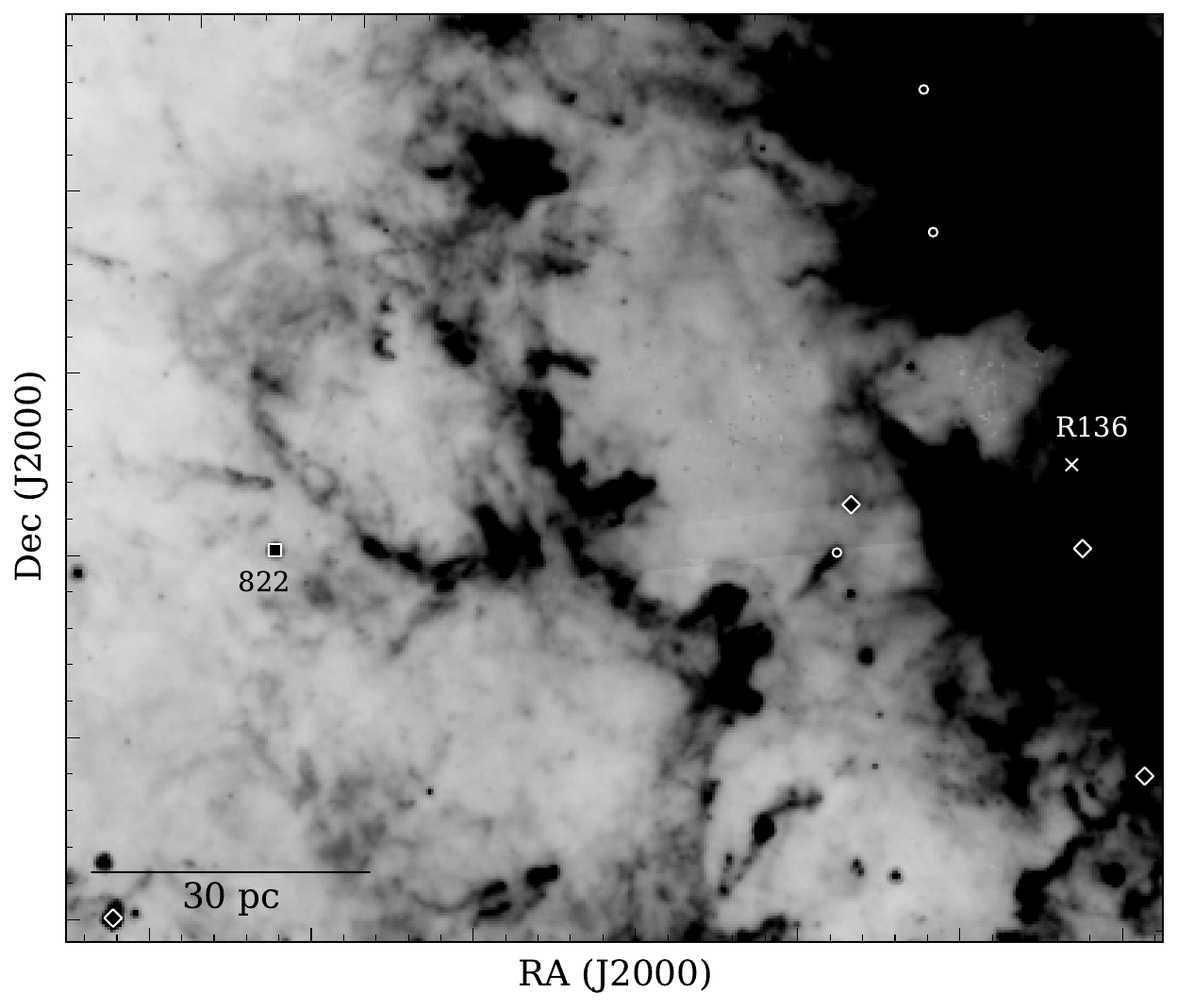}}
\caption{Spitzer 8\,$\mu$m image showing the position of VFTS\,822 (square) with respect to R136 (cross). North is up and east is to the left. Circles and diamonds are known Class I and Class II sources, respectively.}  
\label{}
\end{figure} 
}

\section{Conclusions}

VFTS\,822 is a B8 star with estimated parameters of~$T_{\text {eff}}$ =~12000\,$\pm$\,3000\,K, log\,{\it g}~=~2.6\,$\pm$\,0.4, and log\,$L$~=~4.04\,$\pm$\,0.25\,$L_{\odot}$. Both H$\alpha$ and H$\beta$ are in emission, along with forbidden and permitted metal lines. A UV and infrared excess is found when compared to model colours. It is considered to exhibit the B[e] phenomenon. Therefore, its location in the H-R diagram does not clarify its evolutionary status. Thus, we suggest that:
 
\begin{enumerate}[(i)]
\item VFTS\,822 may be an sgB[e] star, around 35\,Myr, which we consider quite unlikely (see Sec.~3.1). 
\item Alternatively, VFTS\,822 could be a Herbig B[e] PMS star with an age of $\approx$~30000\,years, a mass of 10\,$M_{\odot}$, and a $\dot{M}_{\text{acc}}$ $\sim$~10$^{-4}$\,$M_{\odot}$$yr^{-1}$. 

\end{enumerate}
The lack of evolutionary abundance enrichment and a UV excess favour a PMS rather than post main sequence star. The low log\,{\it g} may be explained as the result of a bloated radius due to the large $\dot{M}_{\text{acc}}$. Becasue no similar PMS stars have been discovered, comparing properties is not feasible at this point. At the measured and predicted $\dot{M}_{\text{acc}}$, VFTS\,822 is expected to have a short PMS lifetime. While the probability of observing a true PMS star in this very small gap of time is serendipitous, our observations suggest that VFTS\,822 is a genuine PMS candidate.

We conclude that if VFTS\,822 is a PMS B[e] star, it is a rare object and worthy of future study. Further data is required to clarify its evolutionary status.   
\\

\end{document}